\documentclass[aps,pra,
    twocolumn,
    longbibliography,
]{revtex4-2}
\usepackage{amsmath,amssymb,amsfonts,amsthm}
\usepackage{bm,mathrsfs,mathtools}
\usepackage{dsfont}
\usepackage{textcomp}
\usepackage{upgreek,textgreek}
\usepackage{physics}
\usepackage{hyperref}
\usepackage{orcidlink}
\usepackage{graphicx}
\usepackage{float}
\usepackage{siunitx}
\usepackage{color}
\usepackage[dvipsnames]{xcolor}

\begin{document}
\title{Stationary electron vortex states in a plasma bubble field}

\author{Hui-Dong Huang\,\orcidlink{0009-0005-8959-8751}}%
\affiliation{\href{https://ror.org/00n9dn158}{Sino-French Institute of Nuclear Engineering and Technology}, \href{https://ror.org/0064kty71}{Sun Yat-Sen University}, Zhuhai 519082, China}

\author{Qi Meng\,\orcidlink{0009-0000-8564-7601}}%
\affiliation{\href{https://ror.org/00n9dn158}{Sino-French Institute of Nuclear Engineering and Technology}, \href{https://ror.org/0064kty71}{Sun Yat-Sen University}, Zhuhai 519082, China}

\author{Zhi-Bin Wang\,\orcidlink{0000-0002-6812-7855}}%
\affiliation{\href{https://ror.org/00n9dn158}{Sino-French Institute of Nuclear Engineering and Technology}, \href{https://ror.org/0064kty71}{Sun Yat-Sen University}, Zhuhai 519082, China}

\author{Liang Lu\,\orcidlink{0000-0002-8497-0738}}%
\affiliation{\href{https://ror.org/00n9dn158}{Sino-French Institute of Nuclear Engineering and Technology}, \href{https://ror.org/0064kty71}{Sun Yat-Sen University}, Zhuhai 519082, China}

\author{Jian Chen\,\orcidlink{0000-0001-9807-489X}}%
\email[Contact author:~]{chenjian5@mail.sysu.edu.cn}
\affiliation{\href{https://ror.org/00n9dn158}{Sino-French Institute of Nuclear Engineering and Technology}, \href{https://ror.org/0064kty71}{Sun Yat-Sen University}, Zhuhai 519082, China}

\author{Li-Ping Zou\,\orcidlink{0000-0001-8976-9171}}%
\email[Contact author:~]{zoulp5@mail.sysu.edu.cn}
\affiliation{\href{https://ror.org/00n9dn158}{Sino-French Institute of Nuclear Engineering and Technology}, \href{https://ror.org/0064kty71}{Sun Yat-Sen University}, Zhuhai 519082, China}

\begin{abstract}
Plasma wakefield accelerators (PWFAs) offer accelerating gradients of 10--100~GV/m and relativistically propagating plasma bubbles capable of confining charged particles. We study the stationary states of a vortex electron at the bubble center by solving the corresponding quasi-relativistic Schrödinger equation. Analytical solutions are obtained with Laguerre--Gaussian transverse modes and Hermite--Gaussian longitudinal envelopes. Comparing the resulting beam parameters with experimentally accessible vortex-electron bundles, we find that the transverse beam waist supported by the plasma bubble is comparable to that achieved by current electron-optical techniques. The longitudinal confinement further provides a favorable parameter regime for stable injection. Our results indicate the feasibility of maintaining localized vortex-electron states in a plasma-bubble wakefield and provide an analytical starting point for investigating their subsequent acceleration and stability.
\end{abstract}
\maketitle

\section{Introduction}

Vortex electrons carrying quantized intrinsic orbital angular momentum (iOAM) \cite{bliokh_2017_theory,lloyd_2017_electron} have emerged as a promising platform at the intersection of beam physics and quantum mechanics. Analogous to optical vortex beams, their wavefunctions can exhibit helical phase structures and phase singularities, giving rise to states with well-defined orbital angular momentum. The iOAM of vortex electrons can reach values of $10^3\hbar$ or higher, far exceeding the $\tfrac{1}{2}\hbar$ spin angular momentum of an electron and providing an additional degree of freedom for controlling angular momentum at the single-particle level \cite{ivanov_2022_promises,zou_2023_recent}. These unique properties have stimulated broad interest in applications ranging from electron microscopy \cite{juchtmans_2015_using} and quantum information \cite{loffler_2023_quantum} to magnetic-field sensing \cite{barrows_2022_3d}, nuclear physics \cite{wu_2022_dynamical}, and particle physics \cite{ivanov_2022_promises,maksimov_2025_diffraction,liu_2025_superkick}.

Early demonstrations of vortex electrons relied primarily on three types of electron-optical elements: spiral phase plates \cite{uchida_2010_generation}, fork gratings and holograms \cite{verbeeck_2010_production,mcmorran_2011_electron,verbeeck_2011_atomic,grillo_2014_generation,grillo_2014_highly}, and spiral zone plates \cite{saitoh_2012_production}. More recently, electromagnetic control of charged-particle wavefunctions has enabled alternative approaches based on artificial magnetic monopoles, or ``magnetic needles'' \cite{beche_2014_magnetic,beche_2017_efficient}, chiral plasmonic near fields \cite{vanacore_2019_ultrafasta}, and electrostatic spiral phase plates \cite{tavabi_2022_generation}, among others \cite{lloyd_2017_electron,bliokh_2017_theory,ivanov_2022_promises,zou_2023_recent}. In these experiments \cite{uchida_2010_generation,verbeeck_2010_production,mcmorran_2011_electron,verbeeck_2011_atomic,saitoh_2012_production,grillo_2014_generation,grillo_2014_highly}, vortex electrons were typically generated using commercial transmission electron microscopes equipped with customized electron-optical elements. The resulting beams could be focused to spot sizes on the order of $1$~\AA \cite{verbeeck_2010_production}, enabling atomic-scale probes of magnetic and electronic properties of materials. The maximum OAM initially demonstrated was approximately $100\hbar$ \cite{mcmorran_2011_electron}, which was subsequently increased to about $200\hbar$ using nanofabricated holographic gratings \cite{grillo_2015_holographic} and, more recently, to the $10^3\hbar$ scale \cite{mafakheri_2017_realization}. Despite these advances, experimentally generated vortex electrons have so far remained at low to intermediate relativistic energies, with ultra-relativistic vortex electrons yet to be demonstrated.

A major obstacle to extending vortex-electron beams to higher energies is the limited radiation hardness and operating energy range of conventional electron-optical elements. A natural strategy is therefore to generate vortex electrons at low or intermediate energies, typically of order $10^2$~keV \cite{ivanov_2022_promises}, and subsequently accelerate them using high-gradient accelerator structures. Recent efforts have explored this approach in conventional radio-frequency linear accelerators, including dedicated studies of vortex-electron dynamics in accelerator components such as magnetic solenoids \cite{ha_2022_bunch,sizykh_2024_transmission,zmaga_2025_radiation,murtazin_2026_photon,karlovets_2026_angular,dyatlov_2026_generation,dyatlov_2026_classical}. However, conventional radio-frequency linacs typically provide accelerating gradients of only about 10--100~MV/m \cite{ha_2022_bunch}. At such gradients, accelerating an electron from rest to 10~MeV corresponds to a proper acceleration time of order $10^1$--$10^2$~ps [see Appendix~\ref{sec6}]. The relatively long interaction time may pose additional challenges for preserving the quantum coherence and spatial structure of vortex-electron wavepackets during acceleration.

Plasma wakefield acceleration (PWFA) offers a fundamentally different route to high-energy vortex electrons. By exploiting collective plasma motion, PWFAs can sustain accelerating gradients of approximately 10--100~GV/m \cite{esarey_2009_physics,tajima_2020_wakefield,lindstrom_2025_beamdriven}, orders of magnitude higher than those of conventional radio-frequency accelerators. Consequently, the proper acceleration time required to reach the MeV energy scale can be reduced to the femtosecond regime [see Appendix~\ref{sec6}], potentially enabling rapid acceleration while mitigating the effects of quantum evolution and decoherence during the acceleration process. In the nonlinear ``bubble'' regime, the plasma wake forms an approximately spherical ion cavity that provides strong, approximately linear electromagnetic focusing fields \cite{lu_2006_nonlinear,esarey_2009_physics}. Such a structure is particularly attractive for vortex electrons because it simultaneously provides transverse confinement and longitudinal acceleration. Nevertheless, the existence and structure of localized quantum vortex eigenstates in a plasma bubble, which constitute the starting point for their subsequent acceleration, remain largely unexplored.

\begin{figure}
\centering
\includegraphics[width=1\linewidth]{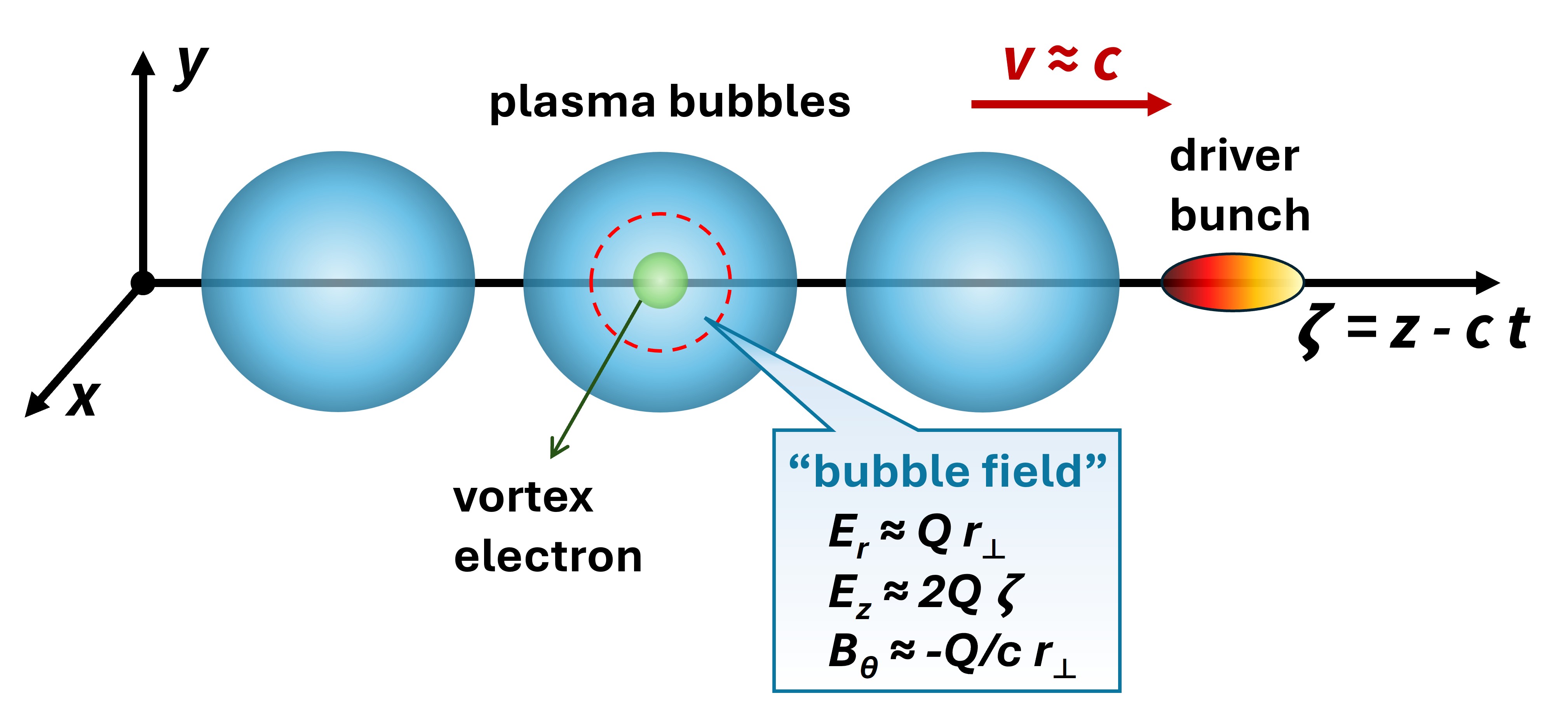}
\caption{Model setup. A vortex-electron wavepacket is localized at the center of a plasma bubble, where the electromagnetic fields vary linearly with position.}
\label{fig:setup}
\end{figure}

In this work, we investigate the stationary states of a vortex electron near the center of a plasma bubble, as sketched in Fig.~\ref{fig:setup}. Starting from the electromagnetic fields of the nonlinear bubble regime \cite{lu_2006_nonlinear,esarey_2009_physics}, we derive the corresponding effective potential and formulate a quasi-relativistic Schrödinger equation for the electron wavefunction. In the bubble co-moving frame, this equation admits analytical stationary solutions with Laguerre--Gaussian (LG) transverse profiles and Hermite--Gaussian (HG) longitudinal envelopes. These solutions provide an explicit description of localized vortex-electron states supported by the plasma bubble and establish their characteristic transverse and longitudinal length scales. By comparing these scales with experimentally accessible vortex-electron bundles, we further assess the feasibility of injecting experimentally generated vortex electrons into the bubble regime.

This paper is organized as follows. Section~\ref{sec2} introduces the plasma-bubble field model. The analytical transverse and longitudinal solutions for the stationary vortex-electron states are derived in Secs.~\ref{sec3} and \ref{sec4}, respectively. The main results are summarized and discussed in Sec.~\ref{sec5}. Additional derivations, including the quasi-relativistic Schrödinger equation and the assessment of higher-order corrections, are provided in the Appendices.

\section{\label{sec2}Plasma Bubble Field}

In a plasma wakefield accelerator, a high-energy driver pulse excites a strong plasma wave and leaves a wake structure behind it. In the so-called blowout, or ``bubble'', regime~\cite{esarey_2009_physics}, the plasma electrons in the wake are completely expelled, forming a nearly electron-free cavity of positive ion charge. This ion cavity, commonly referred to as a plasma bubble, propagates with a velocity approaching the speed of light, $c$. In the co-moving cylindrical coordinates $(r_\perp, \theta, \zeta \equiv z - ct)$, the ion bubble is approximately spherical and remains stable over the propagation length relevant to particle acceleration~\cite{lu_2006_nonlinear,manwani_2025_analysis}.

\begin{figure}
\centering
\includegraphics[width=1\linewidth]{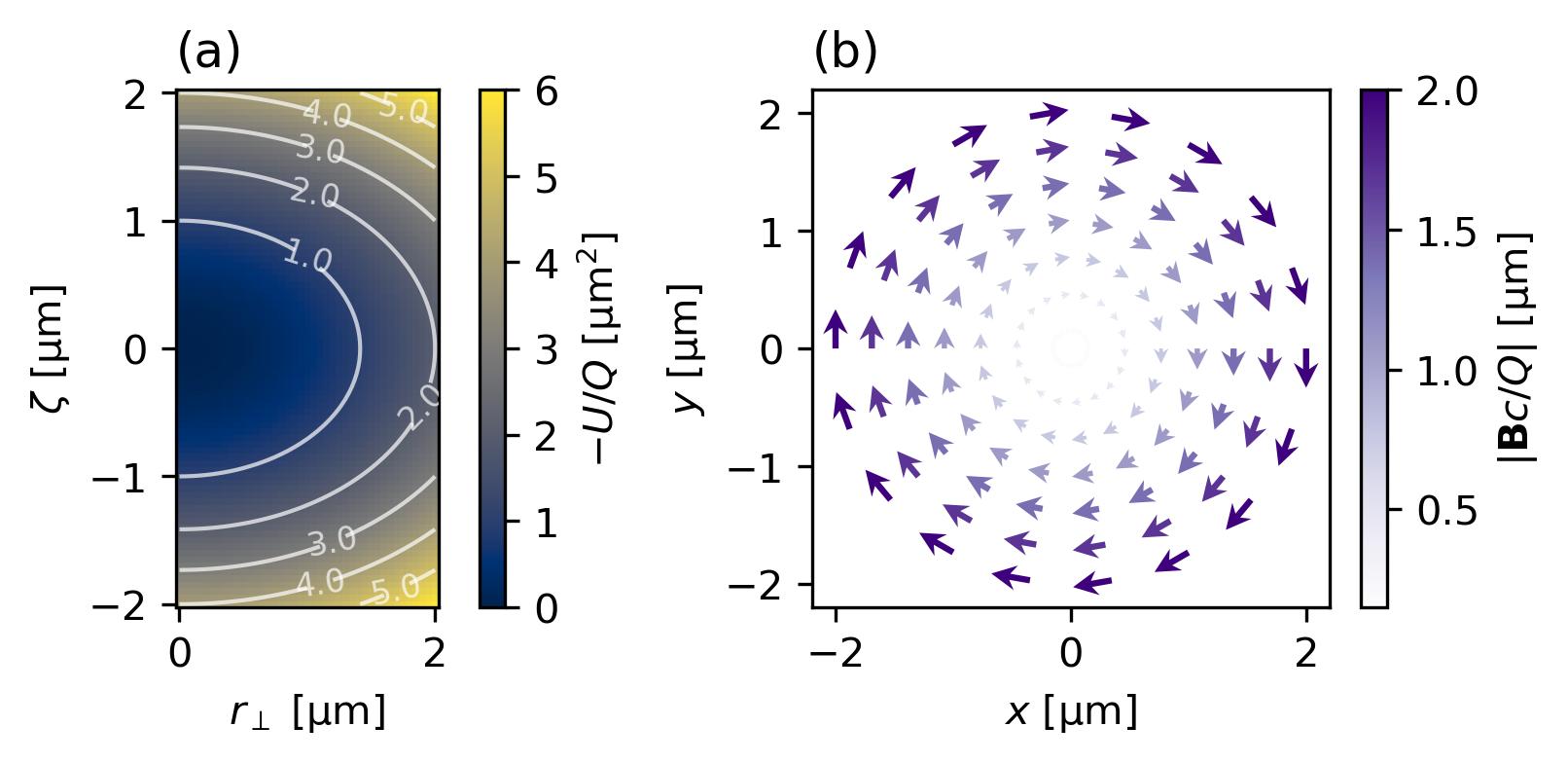}
\caption{The bubble field model formulated by Eq.~\eqref{eq:bubble-field}: (a) scalar wake potential and (b) magnetic field.}
\label{fig:field}
\end{figure}

Near the bubble center, the electromagnetic fields can be approximated by the linearized model:
\begin{equation}\label{eq:bubble-field}
	\begin{aligned}
		&\mathbf{E} = Q\left(r_\perp\,\mathbf{e}_r + 2\zeta\,\mathbf{e}_z\right), \quad
		\mathbf{B} = -\frac{Q}{c} r_\perp\,\mathbf{e}_\theta, \\
		&U = -Q\left(\tfrac{1}{2} r_\perp^2 + \zeta^2\right), \qquad
		\mathbf{A} = \frac{Q}{2c} r_\perp^2\,\mathbf{e}_z,
	\end{aligned}
\end{equation}
where $\mathbf A$ is the vector potential in the Coulomb gauge, $\div \mathbf{A} = 0$. Here, $U$ denotes the corresponding scalar wake potential, and $Q$ characterizes the local field gradient. We have:
\begin{equation}\begin{aligned}
	&\omega_p = \sqrt{\frac{n_0 e^2}{\epsilon_0 m}}, \qquad
	k_p = \frac{\omega_p}{c}, \\
	&E_0 = \frac{m}{e} \omega_p c, \qquad
	Q = \tfrac{1}{4} k_p E_0 = \frac{n_0 e}{4 \epsilon_0},
\end{aligned}\end{equation}
where $n_0$ means the unperturbed plasma density, $e>0$ is elementary charge, $m$ denotes the rest mass of electron unless otherwise specified. The corresponding field configuration is illustrated in Fig.~\ref{fig:field}. For typical plasma densities of $n_0=10^{16}$--$10^{19}\,\mathrm{cm^{-3}}$ used in plasma-based acceleration experiments~\cite{esarey_2009_physics}, the field-gradient coefficient $Q$ falls in the range $Q= 10^{13}$--$10^{16}\,\mathrm{V/m^2}$.

The bubble fields provide strong linear focusing of relativistic electrons. For an electron with charge $-e$ and longitudinal velocity $v_z=\beta_e c$, the transverse Lorentz force is
\begin{equation}
	F_\perp = -e\left(E_r - v_z B_\theta\right) = -eQr_\perp(1+\beta_e).
\end{equation}
Thus, for a highly-relativistic electron with $\beta_e \to 1$, the electric and magnetic contributions add constructively in the focusing force. The resulting linear focusing provides a stable confining potential for the accelerated electron beam~\cite{lu_2006_nonlinear}.

\section{\label{sec3}Transverse Stationary States}

For electrons co-propagating with the plasma wake, $\beta_e \to 1$, the longitudinal momentum is predominantly aligned with the propagation direction. Neglecting spin effects and antiparticle solutions, the longitudinal and transverse dynamics can be decoupled: the longitudinal motion is treated as classical and quasi-stationary, while the transverse motion is described quantum mechanically within a quasi-relativistic Schrödinger framework. The longitudinal velocity enters the transverse dynamics through the Lorentz factor $\gamma=[1-v_z^2/c^2]^{-1/2}$. The transverse envelope of the electron wavepacket can then be described by the effective Hamiltonian
\begin{equation}\label{eq:rel-hamiltonian-2d}
	\hat{H}_\perp = \frac{\hat{p}_\perp^2}{2m\gamma} - q A_z(r_\perp) v_z + \frac{q^2 A_z^2(r_\perp)}{2m\gamma} + q U(r_\perp),
\end{equation}
where $q=-e$ is the electron charge. This effective description can be obtained, for example, by applying a slowly varying envelope approximation to the Klein--Gordon equation [see Appendix~\ref{sec7}] or by taking the appropriate nonrelativistic limit of the three-dimensional Dirac equation. Since the explicit form and interpretation of relativistic vortex solutions of the Dirac equation remain under discussion~\cite{bliokh_2011_relativistic,barnett_2017_relativistic,bialynicki-birula_2017_relativistic,fukushima_2020_mode}, we restrict the present analysis to the quasi-relativistic Schrödinger framework.

For a longitudinal Lorentz boost, the transverse coordinates $(x,y)$, and hence the azimuthal coordinate $\theta$, are invariant. The axial orbital-angular-momentum operator therefore retains the form
\begin{equation}
\hat{L}_z = -i\hbar \partial_\theta.
\end{equation}
Because the transverse Hamiltonian is axially symmetric, $\comm*{\hat{H}}{\hat{L}_z}=0$, and the transverse eigenstates can be chosen as eigenstates of $\hat{L}_z$ with azimuthal dependence
\begin{equation}
\psi(r_\perp,\theta)\propto e^{i\ell\theta},
\qquad
\hat L_z\psi=\ell\hbar\psi,
\end{equation}
where $\ell\in\mathbb Z$. Thus, the plasma bubble field supports vortex states with quantized axial orbital angular momentum $\ell\hbar$. The integer $\ell$ is the winding number of the phase around the vortex core.

Under the paraxial approximation, the term $A_z^2 \propto r_\perp^4$ represents a higher-order perturbation and can be safely neglected [more discussions see Appendix~\ref{sec8}]. Substituting the electromagnetic fields from \eqref{eq:bubble-field} into Eq.~\eqref{eq:rel-hamiltonian-2d}, taking the ultrarelativistic limit $\beta_e\to1$, and truncating the $r_\perp$-dependence up to second order, the transverse Hamiltonian therefore reduces to
\begin{equation}
	\hat{H}_\perp \approx -\frac{\hbar^2}{2m\gamma}\laplacian_\perp - qQ r_\perp^2 .
\end{equation}
For an electron, $q=-e$, and hence the quadratic term is positive since $Q>0$, corresponding to a confining harmonic  potential. The transverse Hamiltonian can consequently be written as that of a two-dimensional isotropic harmonic oscillator with the effective angular frequency
\begin{equation}\label{eq:omega_r}
	\Omega_\perp \equiv \sqrt{\frac{2eQ}{m\gamma}} .
\end{equation}

In cylindrical coordinates, the eigen-functions can be separated as $\psi(r_\perp,\theta) \propto R(r_\perp)\,\mathrm{e}^{\mathrm{i} \ell \theta}$, where the radial function satisfies
\begin{equation}\begin{aligned}
&\hat H_\perp R(r_\perp) = \mathcal E_\perp R(r_\perp) \\
&\hat H_\perp = \left[-\frac{\hbar^2}{2m\gamma}\left(\dv[2]{r_\perp}+\frac{1}{r_\perp}\dv{r_\perp}-\frac{\ell^2}{r_\perp^2}\right)+\frac{1}{2}m\gamma\Omega_\perp^2 r_\perp^2\right],
\end{aligned}\end{equation}
with $\mathcal E_\perp$ denoting the transverse energy eigenvalue. Introducing the characteristic quantum beam waist
\begin{equation}\label{eq:w_r}
w_\perp
\equiv \sqrt{\frac{\hbar}{m\gamma\Omega_\perp}}
= \sqrt{\frac{\hbar}{\sqrt{2m\gamma eQ}}}
\end{equation}
and the dimensionless radial coordinate $\xi \equiv r_\perp / w_\perp$, one has $\mathrm{d}\, r_\perp = w_\perp \mathrm{d}\, \xi$ and $\hbar\Omega_\perp = \hbar^2 / (\gamma m w_\perp^2)$. Dividing the radial equation by $\tfrac{1}{2}\hbar\Omega_\perp$ gives
\begin{equation}\label{eq:radial-schrodinger-adim}
	\left[-\left(\dv[2]{\xi}+\frac{1}{\xi}\dv{\xi}-\frac{\ell^2}{\xi^2}\right)+\xi^2\right] R(\xi) = \lambda R(\xi),
\end{equation}
where $\lambda \equiv 2\mathcal E_\perp / (\hbar\Omega_\perp)$ is the dimensionless radial energy.

The analytical solution can be obtained by examining the asymptotic behavior of $R(\xi)$. For $\xi \to \infty$, Eq.~\eqref{eq:radial-schrodinger-adim} reduces to $\left[-\dv[2]{\xi}+\xi^2\right]R \approx 0$, which gives the normalizable behavior $R(\xi) \sim \exp(-\xi^2/2)$. In the opposite limit, $\xi \to 0$, the dominant terms satisfy $\left[\dv[2]{\xi}+\frac{1}{\xi}\dv{\xi}-\frac{\ell^2}{\xi^2}\right]R \approx 0$, yielding $R(\xi) \sim \xi^{\abs{\ell}}$. These asymptotic forms motivate the ansatz
\begin{equation}
	R(u) = u^{\abs{\ell}/2} \exp(-u/2) f(u),
\end{equation}
where $u \equiv \xi^2$. The derivatives transform according to $\dv{\xi} = 2\xi \dv{u}$ and $\dv[2]{\xi} = 4u \dv[2]{u} + 2 \dv{u}$. Substitution into Eq.~\eqref{eq:radial-schrodinger-adim} gives
\begin{equation}
	\left[4u\dv[2]{u} + 4\dv{u} - \frac{\ell^2}{u} - u + \lambda \right] R(u) = 0 .
\end{equation}
After inserting the ansatz and factoring out the nonvanishing prefactor $\mathrm{e}^{-u/2}u^{\abs{\ell}/2}$, one obtains
\begin{equation}\label{eq:gen-laguerre}
	u\dv[2]{f}{u} + (\abs{\ell} + 1 - u)\dv{f}{u} + \frac{\lambda - 2\abs{\ell} - 2}{4} f = 0 .
\end{equation}
This is the standard generalized Laguerre differential equation,
\begin{equation}
	x y'' + (\alpha + 1 - x) y' + n y = 0,
\end{equation}
whose polynomial solutions are the generalized Laguerre polynomials $L_n^{\alpha}(x)$. Comparison with Eq.~\eqref{eq:gen-laguerre} identifies
\begin{equation}
\alpha = \abs{\ell},\qquad
n = \tfrac{1}{4}(\lambda - 2\abs{\ell} - 2).
\end{equation}

Normalizability requires the Laguerre series to terminate, which restricts $n$ to non-negative integers, $n = 0, 1, 2, \dots$. This resulting transverse energy spectrum is therefore
\begin{equation}
	\mathcal E_\perp = \hbar \Omega_\perp (2n + \abs{\ell} + 1).
\end{equation}
This is the familiar spectrum of a two-dimensional isotropic harmonic oscillator, with $n$ and $\ell$ denoting the radial and azimuthal quantum numbers, respectively.

The corresponding radial function $f(u)$ is given by the generalized Laguerre polynomial $L_n^{\abs{\ell}}(u)$ and, in terms of the physical radial coordinate, takes the Laguerre–Gaussian form
\begin{equation}
	R_{n\ell}(r_\perp) = \mathcal{N}_\perp \frac{r_\perp^{\abs{\ell}}}{w_\perp^{\abs{\ell}}}
	L_n^{\abs{\ell}}\!\left(\frac{r_\perp^2}{w_\perp^2}\right)
	\exp\left(-\frac{r_\perp^2}{2w_\perp^2}\right).
\end{equation}
where
\begin{equation}
	\mathcal{N}_\perp = \frac{1}{w_\perp}\left[\frac{n!}{\pi (n + \abs{\ell})!}\right]^{1/2}.
\end{equation}
The complete transverse eigenstate is thus
\begin{equation}\label{eq:LG}
\psi_{n\ell}(r_\perp, \theta) = R_{n\ell}(r_\perp)\exp(i\ell\theta).
\end{equation}

\begin{figure}[htbp]
\centering
\includegraphics[width=1\linewidth]{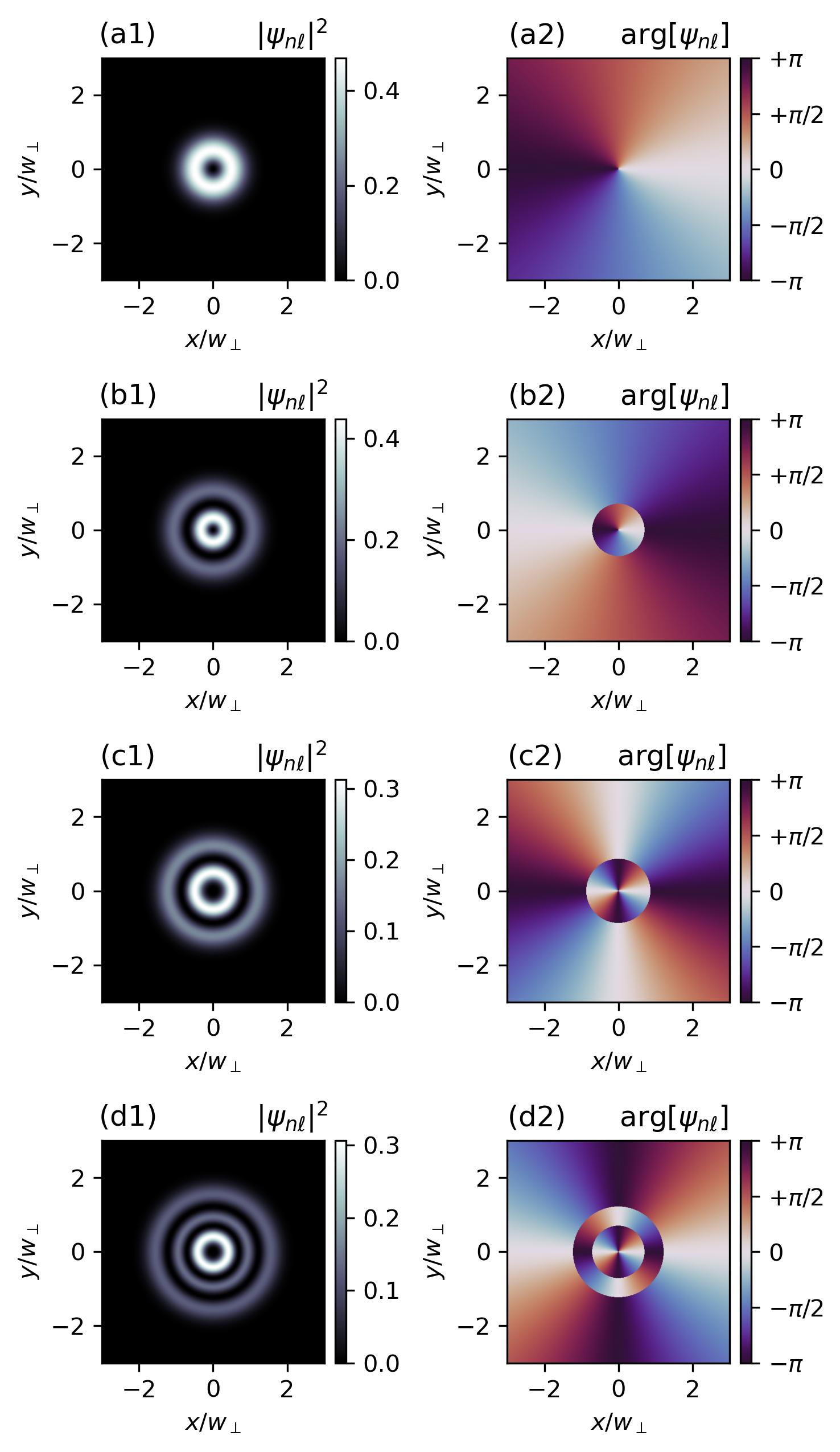}
\caption{Laguerre--Gaussian-type vortex wavefront. Left and right columns depict the transverse distribution of probability density and topological phase respectively. Rows (a--d) present eigenstates $(n,\ell)=(0, +1), (1, +1), (1, +2), (2, +2)$ respectively.}
\label{fig:LG}
\end{figure}

The transverse eigenstates therefore have the same Laguerre--Gaussian structure as those encountered in free-space paraxial optics and other two-dimensional harmonic-confinement problems. Several representative LG-type vortex-eigenstate wavefunctions are showcased in Fig.~\ref{fig:LG}. The radial quantum number $n$ determines the number of radial nodes, while the integer $\ell$ specifies the phase winding around the vortex core and the corresponding axial orbital angular momentum. Near the origin, the radial wavefunction scales as $R_{n\ell}\propto r_\perp^{|\ell|}$, so states with nonzero $\ell$ possess a phase singularity accompanied by a central density minimum. Increasing $|\ell|$ enhances the suppression of probability density $\rho=|\psi_{n\ell}|^2$ near the axis and generally enlarges the characteristic vortex core. The sign of $\ell$ determines the sense of phase winding and the direction of the intrinsic orbital angular momentum, whereas its magnitude determines the winding number and the corresponding angular-momentum magnitude.

\section{\label{sec4}Longitudinal Stationary States}

The longitudinal eigenstates can be obtained by treating the electron motion in the co-moving coordinate $\zeta = z - \beta ct$. In this frame, the longitudinal component of the bubble potential provides a harmonic confining potential, while the relativistic inertia of the electron is enhanced by its forward motion. The resulting longitudinal envelope can therefore be described by an effective one-dimensional harmonic oscillator.

From the bubble-field model in Eq.~\eqref{eq:bubble-field}, the longitudinal scalar wake potential is $U(\zeta)=-Q\zeta^2$. For an electron with $q=-e$, the corresponding longitudinal potential energy reads
\begin{equation}
	V_z(\zeta) = qU(\zeta) = e Q \zeta^2,
\end{equation}
which is equally a confining parabolic potential. In the bubble co-moving frame, the longitudinal dynamics is governed by the response of the electron momentum to a small perturbation of its longitudinal velocity. Starting from the relativistic momentum $p_z = \gamma m v_z$, the corresponding longitudinal inertial mass is
\begin{equation}
	m_z = \dv{p_z}{v_z} = m\left[\gamma + v_z\dv{\gamma}{v_z}\right] = \gamma^3 m .
\end{equation}
Thus, within the quasi-relativistic envelope approximation, the longitudinal Hamiltonian takes the form
\begin{equation}
\hat{H}_z = -\frac{\hbar^2}{2m_z} \dv[2]{\zeta} + V_z,\quad V_z = \frac{1}{2}m_z\Omega_z^2\zeta^2,
\end{equation}
with the effective longitudinal angular frequency
\begin{equation}\label{eq:omega_z}
	\Omega_z \equiv \sqrt{\frac{2eQ}{m\gamma^3}} .
\end{equation}

Using the transverse frequency \eqref{eq:omega_r}, the longitudinal and transverse frequencies are related by $\Omega_z={\Omega_\perp}/\gamma$. Thus, in the ultrarelativistic regime, the longitudinal confinement becomes parametrically weaker than the transverse confinement. This anisotropy originates from the $\gamma^3$ enhancement of the longitudinal inertial mass and is a characteristic feature of relativistic motion in the plasma bubble.

Analogous to the transverse width, the characteristic longitudinal buddle size (longitudinal eigen-extent) $w_z$ is introduced as
\begin{equation}\label{eq:w_z}
	w_z \equiv \sqrt{\frac{\hbar}{m_z \Omega_z}} = \sqrt{\frac{\hbar}{\sqrt{2 m \gamma^3 eQ}}} .
\end{equation}

\begin{figure}[htbp]
\centering
\includegraphics[width=1\linewidth]{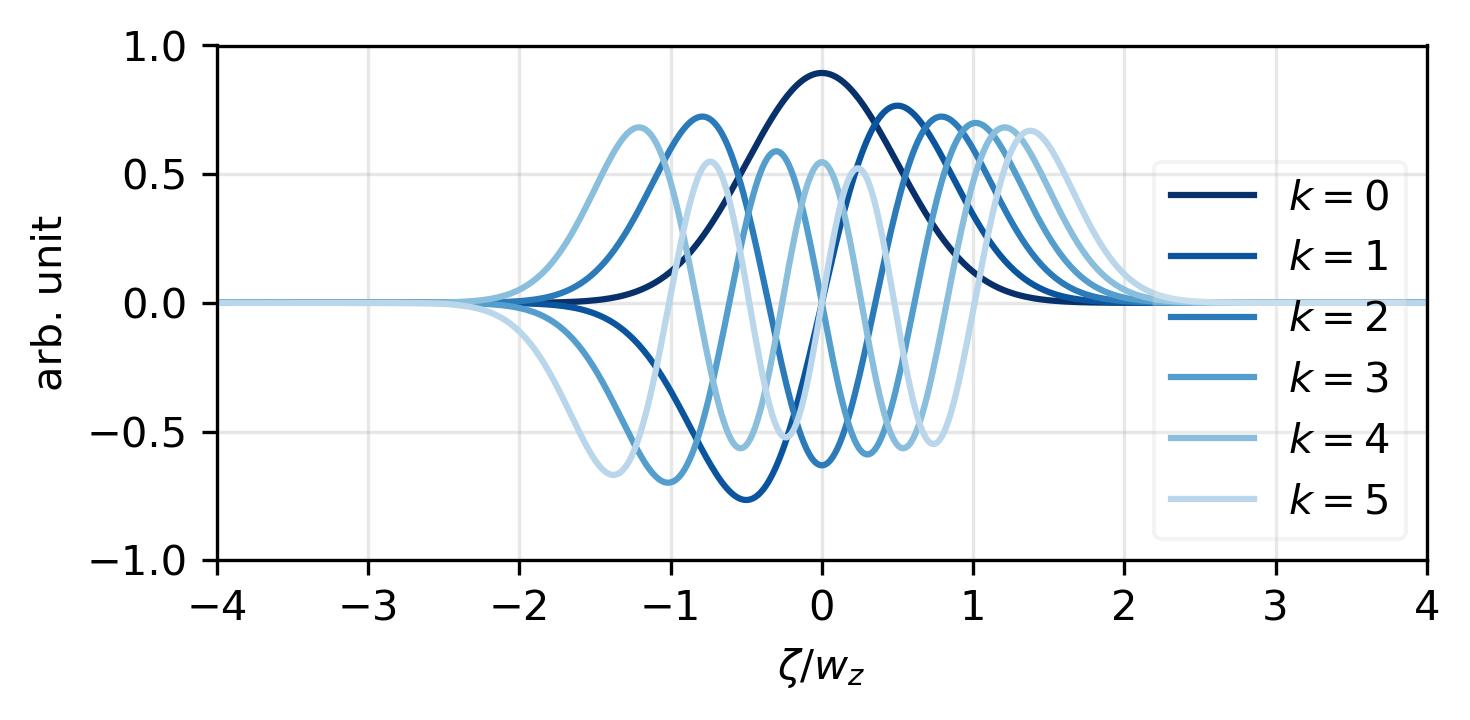}
\caption{Hermite--Gaussian envelopes described by Eq.~\eqref{eq:HG}.}
\label{fig:HG}
\end{figure}

For the longitudinal one-dimensional harmonic oscillator with eigenenergy $\mathcal E_z$,
\begin{equation}
	\left[ -\frac{\hbar^2}{2m\gamma^3} \dv[2]{\zeta} + \frac{1}{2}\left(m\gamma^3 \Omega_z^2\right) \zeta^2 \right] \psi_z(\zeta) = \mathcal E_z \psi_z(\zeta) ,
\end{equation}
its normalized eigenfunctions are the Hermite--Gaussian (HG) modes,
\begin{equation}\label{eq:HG}
	\psi_k(\zeta) = \mathcal{N}_{z}\, H_k\left(\frac{\zeta}{w_z}\right) \exp\left(-\frac{\zeta^2}{2 w_z^2}\right) ,
\end{equation}
where $H_k$ denotes the Hermite polynomial and $k = 0, 1, 2, \dots$ is the longitudinal quantum number, as illustrated in Fig.~\ref{fig:HG}. The integer $k$ determines the number of longitudinal nodes of the envelop. With the convention adopted in Eq.~\eqref{eq:w_z}, the normalization factor is
\begin{equation}
\mathcal N_z = \frac{1}{\pi^{1/4}\sqrt{2^k\, k!\, w_z}}.
\end{equation}
The corresponding longitudinal energy eigenvalues are
\begin{equation}
\mathcal E_z(k) = \left(k + \tfrac{1}{2}\right) \hbar \Omega_z .
\end{equation}

\section{\label{sec5}Summary and Discussion}

The preceding analysis yields an approximate factorized three-dimensional stationary state for a vortex electron confined in the linear focusing field of an ion bubble,
\begin{equation}
\psi_{n\ell k}(r_\perp,\theta,\zeta)
\approx
\psi_k(\zeta)\,
\psi_{n\ell}(r_\perp,\theta).
\end{equation}
The resulting quantum wavepacket has a Hermite--Gaussian (HG) envelope along the propagation direction and a Laguerre--Gaussian (LG) vortex structure in the transverse plane. This analytical solution highlights several important physical features of vortex-electron confinement in the plasma bubble.

\textit{(1) Axial-angular-momentum conservation and vortex-state confinement.}---
Near the bubble center, the electromagnetic field is approximately axially symmetric and produces a quadratic transverse confining potential. Consequently, $[\hat H_\perp,\hat L_z]=0$, so that the transverse eigenstates can be chosen simultaneously as eigenstates of the axial orbital-angular-momentum operator. The topological charge $\ell$ is therefore conserved within the present axisymmetric model, and states belonging to different $\ell$ sectors do not mix. The LG vortex modes thus constitute stationary transverse eigenmodes of the bubble field. This result indicates that the ion-bubble focusing field can provide a stable confinement mechanism for the vortex structure without intrinsically destroying its phase winding.

\textit{(2) Relativistic separation of transverse and longitudinal dynamics.}---
Within the quasi-relativistic approximation, the three-dimensional problem separates into transverse and longitudinal harmonic-oscillator problems. Their characteristic frequencies are related by
\begin{equation}
\Omega_z={\Omega_\perp}/{\gamma}.
\end{equation}
Thus, for $\gamma\gg1$, the longitudinal dynamics evolves on a parametrically slower frequency scale than the transverse motion. This separation of scales provides a useful basis for treating the transverse vortex structure as a rapidly confined mode while the longitudinal envelope evolves more slowly. The result is particularly relevant for injection and transport problems, although the adiabaticity of this separation during acceleration requires further analysis when $\gamma$ varies appreciably along the trajectory.

\textit{(3) Characteristic quantum beam sizes.}---
The transverse and longitudinal confinement scales are characterized by
\begin{equation}
w_\perp
=
\sqrt{\frac{\hbar}{m\gamma\Omega_\perp}},
\qquad
w_z
=
\sqrt{\frac{\hbar}{m_z\Omega_z}}.
\end{equation}
For fixed $\gamma$, both characteristic lengths scale as
\begin{equation}
w_\perp\propto Q^{-1/4},
\qquad
w_z\propto Q^{-1/4}.
\end{equation}
Hence, increasing the plasma field gradient leads to tighter quantum confinement in both transverse and longitudinal directions, while the longitudinal confinement additionally depends strongly on the relativistic factor through $w_z\propto\gamma^{-3/4}$. These scaling relations provide direct criteria for estimating the spatial extent of vortex-electron eigenmodes from the plasma density and electron energy.

\textit{(4) Applicability and limitations.}---
The present analytical treatment relies on several approximations. First, the plasma-bubble fields are linearized about the bubble center, so that higher-order spatial variations are neglected. Second, the quartic contribution arising from the $A_z^2$ term is omitted within the paraxial approximation. Third, the longitudinal motion is treated as a classical, slowly varying background, while the transverse and longitudinal envelopes are quantized within an effective quasi-relativistic Schrödinger description. These approximations are expected to remain valid when the transverse beam size is substantially smaller than the bubble radius and the electron remains highly relativistic and close to the bubble's phase velocity. For larger transverse excursions, strongly nonlinear regions of the bubble, or nonadiabatic acceleration, higher-order field corrections and a fully time-dependent treatment will be required. In particular, the present stationary solutions describe the local eigenmodes at a given relativistic factor $\gamma$ and do not by themselves establish adiabatic preservation of the vortex state during acceleration.

\begin{figure}[htbp]
\centering
\includegraphics[width=1\linewidth]{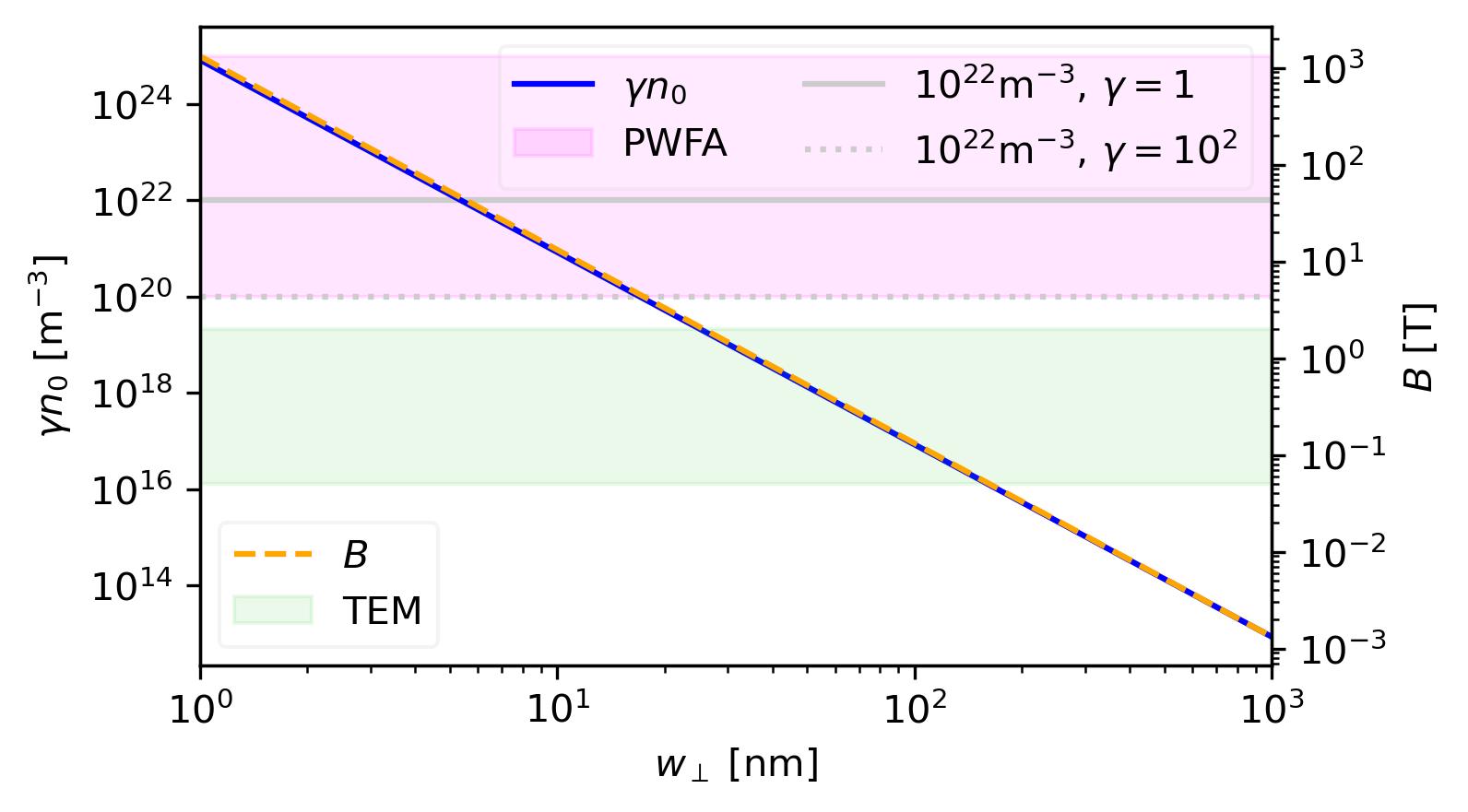}
\caption{Comparison of the characteristic transverse width of vortex-electron eigenmodes in plasma bubbles and magnetic electron-optical systems. The transverse width $w_\perp$ in the plasma-bubble model depends on both the plasma density $n_0$ and the Lorentz factor $\gamma$, while the corresponding magnetic confinement scale in a transmission electron microscope depends on the magnetic field strength $B$. The shaded regions indicate representative parameter ranges, $n_0=\qtyrange{1e22}{1e25}{m^{-3}}$ for PWFAs and $B=\qtyrange{0.05}{2.0}{T}$ for TEM-based magnetic electron optics. The comparable order of the resulting transverse length scales indicates that experimentally generated vortex electrons can, in principle, be transversely matched to the characteristic confinement scale of the plasma bubble.}
\label{fig:wr}
\end{figure}

An important consequence of the analytical solution is that the characteristic transverse confinement scale can be compared directly with experimentally accessible vortex-electron beams. For the field-gradient range $Q=10^{13}$--$10^{16}\,\mathrm{V/m^2}$, relevant to plasma densities of approximately $10^{22}$--$10^{25}\,\mathrm{m^{-3}}$, the transverse eigenmode width falls in the nanometer range, with typical values of order $10^0$--$10^1$~nm for the parameters considered here. The corresponding transverse angular frequency $\Omega_\perp$ lies in the terahertz regime, as shown in Fig.~\ref{fig:wr}. These scales are comparable to the characteristic transverse dimensions of vortex-electron states generated using magnetic electron-optical elements~\cite{verbeeck_2010_production,mcmorran_2011_electron,schattschneider_2014_imaging}, suggesting that the transverse phase-space matching required for injection into a plasma bubble is not intrinsically prohibitive.

For comparison, a magnetic-field-controlled vortex state in a transmission electron microscope provides a useful reference scale. In the experiment of Ref.~\cite{schattschneider_2014_imaging}, a magnetic field of $B=\SI{1.9}{T}$ corresponds to the magnetic length $w_m=\sqrt{2\hbar/(eB)}=\SI{26.3}{nm}$, which is of a comparable order of magnitude as the transverse confinement lengths obtained from the plasma-bubble model [Fig.~\ref{fig:wr}]. The corresponding Larmor frequency is $f_c= eB/(2\pi m)=\SI{53.2}{GHz}$. The transverse oscillation frequencies in the plasma bubble can reach the THz range because the bubble provides a strong electric--magnetic focusing gradient rather than a uniform magnetic field.

The longitudinal scale presents a more restrictive consideration for practical injection. Vortex-electron beams generated in transmission electron microscopes can have longitudinal extents on the micrometer scale~\cite{mcmorran_2011_electron}, whereas the stationary HG modes of the plasma bubble can possess substantially different longitudinal confinement lengths depending on $\gamma$ and $Q$. Efficient injection therefore requires appropriate longitudinal phase-space matching between the incoming vortex-electron wavepacket and the bubble eigenmode. In particular, a significant mismatch in longitudinal width or phase can excite higher-order HG modes and lead to longitudinal breathing during propagation. Preparing the incoming beam close to the corresponding longitudinal eigenmode should consequently provide a smoother transition into the plasma-bubble confinement regime.

In summary, we have derived an analytical description of localized vortex-electron eigenstates in the linear focusing field of a plasma bubble. Within the quasi-relativistic approximation, the three-dimensional problem separates into a transverse two-dimensional harmonic oscillator and a longitudinal one-dimensional harmonic oscillator, yielding a factorized LG--HG stationary state. The transverse mode carries a conserved axial orbital angular momentum $\ell\hbar$, while the characteristic confinement scales are determined by the plasma field gradient and the relativistic factor. The resulting nanometer-scale transverse widths are comparable to those accessible with existing electron-optical techniques, indicating that transverse injection of vortex electrons into a plasma bubble is physically plausible. The present results therefore provide an analytical starting point for studying the injection, acceleration, and dynamical stability of high-energy vortex electrons in plasma wakefields. A quantitative treatment of acceleration-induced mode evolution, higher-order bubble-field corrections, and nonadiabatic effects remains an important direction for future work. A fully time-dependent treatment of vortex-electron acceleration based on the Dirac equation, including fully relativistic quantum dynamics and nonadiabatic mode evolution, will be pursued in future work.

\section*{Acknowledgments}

The work was supported by the National Key R\&D Program of China (No. 2024YFE0109802), the Na-tional Natural Science Foundation of China (Grant No.12175320), and the Guangdong Basic and Applied BasicResearch Foundation (Grant No.2026A1515010999)

\appendix

\section{\label{sec6}Estimating proper acceleration time}

The proper time interval for an electron with initial kinetic energy to gain kinetic energy from an initial value $\mathcal K_i$ to a target value $\mathcal K_f$ can be estimated analytically for a uniform accelerating electric field, $\mathbf{E}=E_0\mathbf{e}_z$, as:
\begin{equation}\label{eq:prop_dur}
\Delta\tau = \frac{\mathcal E_0}{gc}\left[\mathopen{\mathrm{cosh}^{-1}}\left(\frac{\mathcal K_f}{\mathcal E_0}+1\right)-\mathopen{\mathrm{cosh}^{-1}}\left(\frac{\mathcal K_i}{\mathcal E_0}+1\right)\right],
\end{equation}
where $\mathcal E_0 = mc^2$ is the rest energy, and $g=\dv{\mathcal E}{z}$ defines the acceleration gradient. Here, $g=-eE_0$. This expression is obtained straightforward by integrating
\begin{equation}
\mathrm{d}\tau
=\frac{\mathrm{d}t}{\gamma}
=\frac{\mathrm{d}z}{\gamma\beta c}
=\frac{\mathcal E_0}{\sqrt{\mathcal E^2 - \mathcal E_0^2}}\frac{\mathrm{d}z}{c}
=\frac{\mathcal E_0}{\sqrt{\mathcal E^2 - \mathcal E_0^2}}\frac{\mathrm{d}\mathcal E}{gc}
\end{equation}
with total energy $\mathcal E=\mathcal K+\mathcal E_0$ for $\mathcal K=\mathcal K_i\to\mathcal K_f$.

\begin{figure}[htbp]
\centering
\includegraphics[width=1\linewidth]{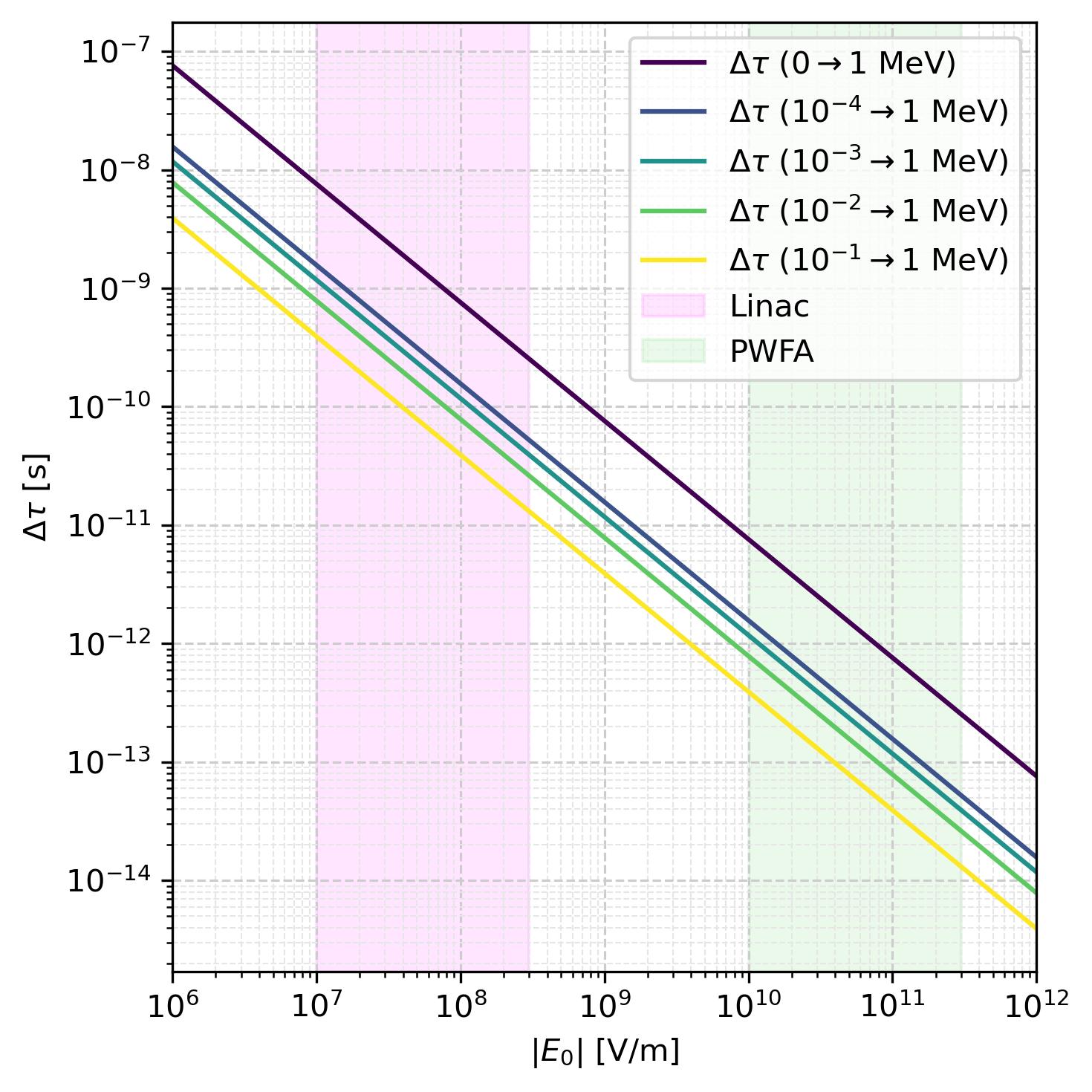}
\caption{Proper acceleration time $\Delta\tau$ as a function of the accelerating electric-field strength $E_0$. Each curve corresponds to an electron accelerated from an initial kinetic energy $\mathcal K_i\in[0,0.1],\mathrm{MeV}$ to the common target energy $\mathcal K_f=\SI{1}{MeV}$. The shaded regions indicate representative accelerating-field ranges for conventional linacs and PWFAs.}
\label{fig:tau_vs_E0}
\end{figure}

Figure~\ref{fig:tau_vs_E0} shows the proper acceleration time calculated from Eq.~\eqref{eq:prop_dur}. For all considered initial energies, increasing the accelerating gradient substantially reduces the proper time required to reach $\mathcal K_f=\SI{1}{MeV}$. In particular, the proper acceleration times associated with conventional linacs and PWFAs differ by several orders of magnitude, illustrating the ultrashort acceleration times enabled by the much larger accelerating gradients available in plasma wakefield acceleration.

\section{\label{sec7}From Klein--Gordon to quasi-relativistic Schr\"odinger equation}

The Klein--Gordon equation for a charged particle in an electromagnetic field,
\begin{equation}
\left[
\frac{1}{c^2}
\left(i\hbar\partial_t-q\Phi\right)^2
-
\left(-i\hbar\nabla-q\bm A\right)^2
-m^2c^2
\right]\Psi=0.
\label{eq:KG}
\end{equation}
For the plasma-bubble configuration considered here,
\begin{equation}
\bm A=A_z\,\bm e_z,
\end{equation}
and the electromagnetic potentials depend on the longitudinal coordinate through the bubble co-moving variable
\begin{equation}
\zeta=z-v_b t,
\label{eq:zeta}
\end{equation}
where $v_b$ is the phase velocity of the bubble.

We write the wavefunction as
\begin{equation}
\Psi(\bm r,t)
=
\exp[
\frac{i}{\hbar}(p_0z-E_0t)
]
\psi(\bm r_\perp,\zeta,t),
\label{eq:carrier}
\end{equation}
where
\begin{equation}
E_0^2=p_0^2c^2+m^2c^4.
\label{eq:dispersion}
\end{equation}
The carrier velocity is
\begin{equation}
v_0=\frac{\partial E_0}{\partial p_0}
=\frac{p_0c^2}{E_0}.
\end{equation}
For a state synchronized with the plasma bubble, we choose
\begin{equation}
v_0=v_b.
\label{eq:synchronism}
\end{equation}
Consequently,
\begin{equation}
E_0=\gamma_bmc^2,
\quad
p_0=\gamma_bmv_b,
\quad
\gamma_b=\frac{1}{\sqrt{1-\beta_b^2}},
\quad
\beta_b=\frac{v_b}{c}.
\end{equation}

Since $\zeta=z-v_bt$, the derivatives acting on the slowly varying envelope satisfy
\begin{equation}
\partial_z\rightarrow\partial_\zeta,
\qquad
\partial_t\rightarrow\partial_t-v_b\partial_\zeta.
\end{equation}
It is convenient to introduce
\begin{equation}\label{eq:Dt}
\mathcal D_t
=
i\hbar\partial_t
-i\hbar v_b\partial_\zeta
-q\Phi,
\end{equation}
and
\begin{equation}\label{eq:Dz}
\mathcal D_z
=
-i\hbar\partial_\zeta-qA_z.
\end{equation}
Then
\begin{equation}
(i\hbar\partial_t-q\Phi)\Psi
=
e^{i(p_0z-E_0t)/\hbar}
(E_0+\mathcal D_t)\psi,
\end{equation}
and
\begin{equation}
(-i\hbar\partial_z-qA_z)\Psi
=
e^{i(p_0z-E_0t)/\hbar}
(p_0+\mathcal D_z)\psi.
\end{equation}

Substituting these expressions into Eq.~\eqref{eq:KG}, and using Eq.~\eqref{eq:dispersion}, gives
\begin{equation}
\left[
\frac{2E_0}{c^2}\mathcal D_t
-2p_0\mathcal D_z
+
\frac{\mathcal D_t^2}{c^2}
-\mathcal D_z^2
+
\hbar^2\nabla_\perp^2
\right]\psi=0.
\label{eq:exact-envelope}
\end{equation}

We consider states localized near the bubble center and centered around the
synchronous longitudinal momentum $p_0$. The envelope therefore describes
small deviations from the reference trajectory in both coordinate and
momentum space. In particular, the transverse momenta, the longitudinal
momentum deviation, and the transverse and longitudinal coordinates are
treated as small quantities.

For the linearized bubble model, the electromagnetic potentials scale as
\begin{equation}
\Phi \propto r_\perp^2+2\zeta^2,
\qquad
A_z \propto r_\perp^2.
\end{equation}
Thus, the field-induced potential energies are already second order in the
deviations from the reference trajectory. We consequently retain the
leading kinetic and potential contributions that determine the local
quadratic Hamiltonian, including
\begin{equation}
p_\perp^2,\qquad
\delta p_z^2,\qquad
q\Phi,\qquad
qv_b A_z,
\end{equation}
while neglecting terms that are higher order in the envelope dynamics, such
as
\begin{equation}
\partial_t^2\psi,\quad
\partial_t\partial_\zeta\psi,\quad
\Phi\,\partial_\zeta\psi,\quad
A_z\,\partial_\zeta\psi,\quad
\Phi^2,\quad
A_z^2.
\label{eq:neglected}
\end{equation}

It is important that this truncation is not the conventional paraxial
slowly varying envelope approximation in which the longitudinal dispersion
term $\partial_\zeta^2\psi$ is discarded. Here we retain the longitudinal
quantum dispersion explicitly. Indeed, the terms linear in the longitudinal
envelope derivative in Eq.~\eqref{eq:exact-envelope} are
\begin{equation}
2i\hbar
\left(
p_0-\frac{E_0v_b}{c^2}
\right)\partial_\zeta\psi.
\end{equation}
For the synchronous carrier, $v_0=v_b$, one has
\begin{equation}
p_0=\frac{E_0v_b}{c^2},
\end{equation}
and this first-order longitudinal term vanishes identically. Therefore,
there is no carrier-scale longitudinal drift in the co-moving coordinate
$\zeta$. The leading nonvanishing longitudinal kinetic contribution is then
the second-order dispersion term $\partial_\zeta^2\psi$, which must be
retained in order to describe longitudinal quantum spreading and to obtain
a finite longitudinal effective mass.

Accordingly, the approximation employed here may be viewed as a
slowly varying envelope expansion with retained longitudinal quantum
dispersion. The envelope is assumed to vary slowly compared with the
carrier scales, while the second-order longitudinal derivative is retained
because it is the leading nonvanishing term governing the longitudinal
dynamics in the synchronous frame.

The terms in Eq.~\eqref{eq:exact-envelope} that are linear in the derivatives are
\begin{equation}
\frac{2E_0}{c^2}
\left(
i\hbar\partial_t
-i\hbar v_b\partial_\zeta
-q\Phi
\right)
-2p_0
\left(
-i\hbar\partial_\zeta-qA_z
\right).
\end{equation}
The remaining first-order terms are therefore
\begin{equation}
\frac{2E_0}{c^2}i\hbar\partial_t
-
\frac{2E_0}{c^2}q\Phi
+
2p_0qA_z.
\label{eq:firstorder}
\end{equation}

Under the above truncation, the surviving second-order longitudinal
derivative originates from the combination
\begin{equation}
\frac{\mathcal D_t^2}{c^2}-\mathcal D_z^2.
\end{equation}
Using Eqs.~\eqref{eq:Dt} and \eqref{eq:Dz} for this contribution, one obtains
\begin{equation}
\frac{\mathcal D_t^2}{c^2}
\simeq
-\hbar^2\frac{v_b^2}{c^2}\partial_\zeta^2,
\qquad
\mathcal D_z^2
\simeq
-\hbar^2\partial_\zeta^2.
\end{equation}
Hence
\begin{equation}\label{eq:longitudinal-second}
\frac{\mathcal D_t^2}{c^2}-\mathcal D_z^2
=
\hbar^2
\left(
1-\frac{v_b^2}{c^2}
\right)\partial_\zeta^2
=
\frac{\hbar^2}{\gamma_b^2}\partial_\zeta^2.
\end{equation}

Substituting Eqs.~\eqref{eq:firstorder} and \eqref{eq:longitudinal-second} into Eq.~\eqref{eq:exact-envelope} yields
\begin{equation}
\frac{2E_0}{c^2}i\hbar\partial_t\psi
-
\frac{2E_0}{c^2}q\Phi\psi
+
2p_0qA_z\psi
+
\frac{\hbar^2}{\gamma_b^2}\partial_\zeta^2\psi
+
\hbar^2\nabla_\perp^2\psi
=0.
\end{equation}
Using
\begin{equation}
E_0=\gamma_bmc^2,
\qquad
\frac{p_0c^2}{E_0}=v_b,
\end{equation}
we obtain the effective quasi-relativistic Schr\"odinger equation
\begin{equation}
i\hbar\partial_t\psi
=
\left[
-\frac{\hbar^2}{2m\gamma_b}\nabla_\perp^2
-\frac{\hbar^2}{2m\gamma_b^3}\partial_\zeta^2
+
q(\Phi-v_bA_z)
\right]\psi.
\label{eq:effective-Sch}
\end{equation}

Thus, the transverse and longitudinal effective inertial masses emerge directly from the Klein--Gordon equation:
\begin{equation}
m_\perp=\gamma_bm,
\qquad
m_\parallel=\gamma_b^3m.
\end{equation}

For the linearized bubble field,
\begin{equation}
\Phi
=
-Q\left(
\frac12r_\perp^2+\zeta^2
\right),
\qquad
A_z=
\frac{Q}{2c}r_\perp^2.
\end{equation}
For an electron, $q=-e$, and therefore
\begin{align}
q(\Phi-v_bA_z)
&=
eQ\zeta^2
+
\frac{eQ}{2}(1+\beta_b)r_\perp^2.
\label{eq:effective-potential}
\end{align}
Equation~\eqref{eq:effective-Sch} becomes
\begin{equation}
i\hbar\partial_t\psi
=
\left[
-\frac{\hbar^2}{2m\gamma_b}\nabla_\perp^2
+
\frac{eQ}{2}(1+\beta_b)r_\perp^2
-\frac{\hbar^2}{2m\gamma_b^3}\partial_\zeta^2
+
eQ\zeta^2
\right]\psi.
\label{eq:bubble-effective}
\end{equation}

The Hamiltonian separates naturally as
\begin{equation}
H_{\rm eff}=H_\perp+H_z,
\end{equation}
with
\begin{equation}
H_\perp
=
-\frac{\hbar^2}{2m\gamma_b}\nabla_\perp^2
+
\frac12m\gamma_b\Omega_\perp^2r_\perp^2,
\end{equation}
and
\begin{equation}
H_z
=
-\frac{\hbar^2}{2m\gamma_b^3}\partial_\zeta^2
+
\frac12m\gamma_b^3\Omega_z^2\zeta^2.
\end{equation}

\section{\label{sec8}Quartic fine structure}

The unperturbed transverse spectrum depends only on the shell index $N=2n+|\ell|$, and is therefore degenerate with respect to different pairs $(n,|\ell|)$ sharing the same $N$. The leading correction neglected in the quadratic transverse Hamiltonian arises from the $A_z^2$ term in the quasi-relativistic Hamiltonian. For the bubble-field model, this contribution is
\begin{equation}\label{eq:quartic-perturbation}
	\delta\hat H_4=\frac{e^2Q^2}{8\gamma mc^2}r_\perp^4 .
\end{equation}
The perturbation is rotationally symmetric and therefore commutes with $\hat L_z$; consequently, it cannot couple states with different $\ell$. It does, however, distinguish states with different radial structures within the same $N$ shell.

Using the normalized Laguerre--Gaussian eigenmodes obtained in Sec.~\ref{sec3}, the required fourth-order radial moment is
\begin{equation}\label{eq:r4-expectation}
\begin{aligned}
&\ev{r_\perp^4}_{n\ell}=w_\perp^4 F_{n\ell},\\
&F_{n\ell}=6n(n+|\ell|+1)+(|\ell|+1)(|\ell|+2).
\end{aligned}
\end{equation}
First-order perturbation theory then gives the quartic energy correction
\begin{equation}\label{eq:quartic-shift}
	\Delta\mathcal E_{n\ell}=\frac{e^2Q^2w_\perp^4}{8\gamma mc^2}F_{n\ell}.
\end{equation}

To assess the magnitude of this correction, we use the relation between the field gradient and the transverse confinement frequency, $eQ=\gamma m\Omega_\perp^2/(1+\beta_e)$, which follows from the quadratic transverse Hamiltonian. Combining this relation with Eq.~\eqref{eq:w_r}, yields the dimensionless correction
\begin{equation}\label{eq:quartic-small-parameter}
	\frac{\Delta\mathcal E_{n\ell}}{\hbar\Omega_\perp}
	=\frac{F_{n\ell}}{8(1+\beta_e)^2}
	\frac{\hbar\Omega_\perp}{\gamma mc^2}
	\xrightarrow{\beta_e\to1}
	\frac{F_{n\ell}}{32}\frac{\hbar\Omega_\perp}{\gamma mc^2}.
\end{equation}

Equation~\eqref{eq:quartic-small-parameter} shows that the validity of the quadratic approximation requires more than the geometric condition that the transverse wavepacket remain well within the bubble radius. For each significantly populated mode, the perturbative correction should also satisfy $\Delta\mathcal E_{n\ell}\ll\hbar\Omega_\perp$. The quartic term preserves the axial-angular-momentum quantum number $\ell$ and therefore leaves the time-reversal-related pair $\pm\ell$ degenerate. At the same time, it lifts the accidental degeneracy between states with different radial structures but the same shell index $N$. For example, within the $N=2$ shell, the states $(n,|\ell|)=(1,0)$ and $(0,2)$ have $F=14$ and $12$, and consequently acquire different first-order energy shifts.

For the plasma densities and relativistic injection energies considered in this work, the dimensionless parameter in Eq.~\eqref{eq:quartic-small-parameter} is expected to remain very small. The quartic contribution therefore produces only a negligible fine-structure correction to the transverse spectrum. Its main significance is not as an independently observable spectral feature, but rather as a quantitative verification that the quadratic Hamiltonian provides a controlled description of the stationary vortex states in the parameter regime of interest.

\bibliography{reference.bib}

@article{verbeeck_2010_production,
title = {Production and Application of Electron Vortex Beams},
author = {Verbeeck, J. and Tian, H. and Schattschneider, P.},
year = 2010,
month = sep,
journal = {Nature},
volume = {467},
number = {7313},
pages = {301--304},
publisher = {Nature Publishing Group},
doi = {10.1038/nature09366},
copyright = {2010 Springer Nature Limited},
lccn = {1}
}

@article{uchida_2010_generation,
	title = {Generation of Electron Beams Carrying Orbital Angular Momentum},
	author = {Uchida, Masaya and Tonomura, Akira},
	year = 2010,
	month = apr,
	journal = {Nature},
	volume = {464},
	number = {7289},
	pages = {737--739},
	doi = {10.1038/nature08904},
	copyright = {http://www.springer.com/tdm}
}

@article{verbeeck_2011_atomic,
	title = {Atomic Scale Electron Vortices for Nanoresearch},
	author = {Verbeeck, J. and Schattschneider, P. and Lazar, S. and {St{\"o}ger-Pollach}, M. and L{\"o}ffler, S. and {Steiger-Thirsfeld}, A. and Van Tendeloo, G.},
	year = 2011,
	month = nov,
	journal = {Applied Physics Letters},
	volume = {99},
	number = {20},
	pages = {203109},
	doi = {10.1063/1.3662012}
}

@article{mcmorran_2011_electron,
	title = {Electron {{Vortex Beams}} with {{High Quanta}} of {{Orbital Angular Momentum}}},
	author = {McMorran, Benjamin J. and Agrawal, Amit and Anderson, Ian M. and Herzing, Andrew A. and Lezec, Henri J. and McClelland, Jabez J. and Unguris, John},
	year = 2011,
	month = jan,
	journal = {Science},
	volume = {331},
	number = {6014},
	pages = {192--195},
	publisher = {American Association for the Advancement of Science},
	doi = {10.1126/science.1198804},
	copyright = {Copyright \copyright{} 2011, American Association for the Advancement of Science},
	lccn = {1}
}

@article{saitoh_2012_production,
	title = {Production of Electron Vortex Beams Carrying Large Orbital Angular Momentum Using Spiral Zone Plates},
	author = {Saitoh, K. and Hasegawa, Y. and Tanaka, N. and Uchida, M.},
	year = 2012,
	month = mar,
	journal = {Journal of Electron Microscopy},
	pages = {dfs036},
	doi = {10.1093/jmicro/dfs036}
}

@article{grillo_2014_generation,
title = {Generation of Nondiffracting Electron Bessel Beams},
author = {Grillo, Vincenzo and Karimi, Ebrahim and Gazzadi, Gian Carlo and Frabboni, Stefano and Dennis, Mark R. and Boyd, Robert W.},
year = 2014,
month = jan,
journal = {Physical Review X},
volume = {4},
number = {1},
pages = {011013},
doi = {10.1103/PhysRevX.4.011013},
copyright = {http://creativecommons.org/licenses/by/3.0/}
}

@article{grillo_2014_highly,
	title = {Highly Efficient Electron Vortex Beams Generated by Nanofabricated Phase Holograms},
	author = {Grillo, Vincenzo and Carlo Gazzadi, Gian and Karimi, Ebrahim and Mafakheri, Erfan and Boyd, Robert W. and Frabboni, Stefano},
	year = 2014,
	month = jan,
	journal = {Applied Physics Letters},
	volume = {104},
	number = {4},
	pages = {043109},
	doi = {10.1063/1.4863564}
}

@article{beche_2014_magnetic,
	title = {Magnetic Monopole Field Exposed by Electrons},
	author = {B{\'e}ch{\'e}, Armand and Van Boxem, Ruben and Van Tendeloo, Gustaaf and Verbeeck, Jo},
	year = 2014,
	month = jan,
	journal = {Nature Physics},
	volume = {10},
	number = {1},
	pages = {26--29},
	doi = {10.1038/nphys2816}
}

@article{schattschneider_2014_imaging,
title = {Imaging the Dynamics of Free-Electron Landau States},
author = {Schattschneider, P. and Schachinger, Th and {St{\"o}ger-Pollach}, M. and L{\"o}ffler, S. and {Steiger-Thirsfeld}, A. and Bliokh, K. Y. and Nori, Franco},
year = 2014,
month = aug,
journal = {Nature Communications},
volume = {5},
number = {1},
pages = {4586},
publisher = {Nature Publishing Group},
doi = {10.1038/ncomms5586},
copyright = {2014 The Author(s)}
}

@article{grillo_2015_holographic,
	title = {Holographic Generation of Highly Twisted Electron Beams},
	author = {Grillo, Vincenzo and Gazzadi, Gian Carlo and Mafakheri, Erfan and Frabboni, Stefano and Karimi, Ebrahim and Boyd, Robert W.},
	year = 2015,
	month = jan,
	journal = {Physical Review Letters},
	volume = {114},
	number = {3},
	pages = {034801},
	doi = {10.1103/PhysRevLett.114.034801},
	copyright = {http://link.aps.org/licenses/aps-default-license}
}

@article{mafakheri_2017_realization,
	title = {Realization of Electron Vortices with Large Orbital Angular Momentum Using Miniature Holograms Fabricated by Electron Beam Lithography},
	author = {Mafakheri, E. and Tavabi, A. H. and Lu, P.-H. and Balboni, R. and Venturi, F. and Menozzi, C. and Gazzadi, G. C. and Frabboni, S. and Sit, A. and {Dunin-Borkowski}, R. E. and Karimi, E. and Grillo, V.},
	year = 2017,
	month = feb,
	journal = {Applied Physics Letters},
	volume = {110},
	number = {9},
	pages = {093113},
	doi = {10.1063/1.4977879}
}

@article{beche_2017_efficient,
	title = {Efficient Creation of Electron Vortex Beams for High Resolution {{STEM}} Imaging},
	author = {B{\'e}ch{\'e}, A. and Juchtmans, R. and Verbeeck, J.},
	year = 2017,
	month = jul,
	journal = {Ultramicroscopy},
	volume = {178},
	pages = {12--19},
	doi = {10.1016/j.ultramic.2016.05.006}
}

@article{vanacore_2019_ultrafasta,
	title = {Ultrafast Generation and Control of an Electron Vortex Beam via Chiral Plasmonic near Fields},
	author = {Vanacore, G. M. and Berruto, G. and Madan, I. and Pomarico, E. and Biagioni, P. and Lamb, R. J. and McGrouther, D. and Reinhardt, O. and Kaminer, I. and Barwick, B. and Larocque, H. and Grillo, V. and Karimi, E. and Garc{\'i}a De Abajo, F. J. and Carbone, F.},
	year = 2019,
	month = jun,
	journal = {Nature Materials},
	volume = {18},
	number = {6},
	pages = {573--579},
	doi = {10.1038/s41563-019-0336-1}
}

@article{barrows_2022_3d,
	title = {{{3D}} Magnetic Imaging Using Electron Vortex Beam Microscopy},
	author = {Barrows, Frank and {Petford-Long}, Amanda K. and Phatak, Charudatta},
	year = 2022,
	month = dec,
	journal = {Communications Physics},
	volume = {5},
	number = {1},
	pages = {1--11},
	publisher = {Nature Publishing Group},
	doi = {10.1038/s42005-022-01082-z},
	copyright = {2022 UChicago Argonne, LLC, Operator of Argonne National Laboratory},
	lccn = {1}
}

@article{tavabi_2022_generation,
	title = {Generation of Electron Vortex Beams with over 1000 Orbital Angular Momentum Quanta Using a Tunable Electrostatic Spiral Phase Plate},
	author = {Tavabi, A. H. and Rosi, P. and Roncaglia, A. and Rotunno, E. and Beleggia, M. and Lu, P.-H. and Belsito, L. and Pozzi, G. and Frabboni, S. and Tiemeijer, P. and {Dunin-Borkowski}, R. E. and Grillo, V.},
	year = 2022,
	month = aug,
	journal = {Applied Physics Letters},
	volume = {121},
	number = {7},
	pages = {073506},
	doi = {10.1063/5.0093411}
}

@article{juchtmans_2015_using,
	title = {Using Electron Vortex Beams to Determine Chirality of Crystals in Transmission Electron Microscopy},
	author = {Juchtmans, Roeland and B{\'e}ch{\'e}, Armand and Abakumov, Artem and Batuk, Maria and Verbeeck, Jo},
	year = 2015,
	month = mar,
	journal = {Physical Review B},
	volume = {91},
	number = {9},
	pages = {094112},
	doi = {10.1103/PhysRevB.91.094112},
	copyright = {http://link.aps.org/licenses/aps-default-license}
}

@article{wu_2022_dynamical,
	title = {Dynamical Control of Nuclear Isomer Depletion via Electron Vortex Beams},
	author = {Wu, Yuanbin and Gargiulo, Simone and Carbone, Fabrizio and Keitel, Christoph H. and P{\'a}lffy, Adriana},
	year = 2022,
	month = apr,
	journal = {Physical Review Letters},
	volume = {128},
	number = {16},
	pages = {162501},
	publisher = {American Physical Society},
	doi = {10.1103/PhysRevLett.128.162501}
}

@article{loffler_2023_quantum,
	title = {A Quantum Logic Gate for Free Electrons},
	author = {L{\"o}ffler, Stefan and Schachinger, Thomas and Hartel, Peter and Lu, Peng-Han and {Dunin-Borkowski}, Rafal E. and Obermair, Martin and Dries, Manuel and Gerthsen, Dagmar and Schattschneider, Peter},
	year = 2023,
	month = jul,
	journal = {Quantum},
	volume = {7},
	pages = {1050},
	doi = {10.22331/q-2023-07-11-1050}
}

@article{bliokh_2017_theory,
	title = {Theory and Applications of Free-Electron Vortex States},
	author = {Bliokh, K. Y. and Ivanov, I. P. and Guzzinati, G. and Clark, L. and Van Boxem, R. and B{\'e}ch{\'e}, A. and Juchtmans, R. and Alonso, M. A. and Schattschneider, P. and Nori, F. and Verbeeck, J.},
	year = 2017,
	month = may,
	journal = {Physics Reports},
	volume = {690},
	pages = {1--70},
	doi = {10.1016/j.physrep.2017.05.006}
}

@article{lloyd_2017_electron,
	title = {Electron Vortices: {{Beams}} with Orbital Angular Momentum},
	shorttitle = {Electron Vortices},
	author = {Lloyd, S. M. and Babiker, M. and Thirunavukkarasu, G. and Yuan, J.},
	year = 2017,
	month = aug,
	journal = {Reviews of Modern Physics},
	volume = {89},
	number = {3},
	pages = {035004},
	publisher = {American Physical Society},
	doi = {10.1103/RevModPhys.89.035004},
	lccn = {1}
}

@article{ivanov_2022_promises,
title = {Promises and Challenges of High-Energy Vortex States Collisions},
author = {Ivanov, Igor P.},
year = 2022,
month = nov,
journal = {Progress in Particle and Nuclear Physics},
volume = {127},
pages = {103987},
doi = {10.1016/j.ppnp.2022.103987},
lccn = {1}
}

@article{zou_2023_recent,
	title = {Recent Progress in the Physics of Twisted Particles},
	author = {Zou, Liping and Zhang, Pengming and Silenko, Alexander J. and Lu, Liang},
	year = 2023,
	month = may,
	journal = {The Innovation},
	volume = {4},
	number = {3},
	pages = {100432},
	doi = {10.1016/j.xinn.2023.100432}
}

@article{bliokh_2011_relativistic,
	title = {Relativistic Electron Vortex Beams: {{Angular}} Momentum and Spin-Orbit Interaction},
	shorttitle = {Relativistic {{Electron Vortex Beams}}},
	author = {Bliokh, Konstantin Y. and Dennis, Mark R. and Nori, Franco},
	year = 2011,
	month = oct,
	journal = {Physical Review Letters},
	volume = {107},
	number = {17},
	pages = {174802},
	doi = {10.1103/PhysRevLett.107.174802},
	lccn = {1}
}

@article{liu_2025_superkick,
	title = {Superkick Effect in Vortex Scattering: {{M\o ller}} Scattering},
	shorttitle = {Superkick Effect in Vortex Scattering},
	author = {Liu, Shiyu and Liu, Bei and Ivanov, Igor P. and Ji, Liangliang},
	year = 2025,
	month = sep,
	journal = {Physical Review A},
	volume = {112},
	number = {3},
	pages = {032814},
	doi = {10.1103/qrfd-9356}
}

@article{barnett_2017_relativistic,
ids = {barnett2017RelativisticElectronVortices},
title = {Relativistic Electron Vortices},
author = {Barnett, Stephen M.},
year = 2017,
month = mar,
journal = {Physical Review Letters},
volume = {118},
number = {11},
pages = {114802},
publisher = {American Physical Society},
doi = {10.1103/PhysRevLett.118.114802}
}

@article{bialynicki-birula_2017_relativistic,
	title = {Relativistic {{Electron Wave Packets Carrying Angular Momentum}}},
	author = {{Bialynicki-Birula}, Iwo and {Bialynicka-Birula}, Zofia},
	year = 2017,
	month = mar,
	journal = {Physical Review Letters},
	volume = {118},
	number = {11},
	pages = {114801},
	publisher = {American Physical Society},
	doi = {10.1103/PhysRevLett.118.114801}
}

@article{fukushima_2020_mode,
	title = {Mode Decomposed Chiral Magnetic Effect and Rotating Fermions},
	author = {Fukushima, Kenji and Shimazaki, Takuya and Wang, Lingxiao},
	year = 2020,
	month = jul,
	journal = {Physical Review D},
	volume = {102},
	number = {1},
	pages = {014045},
	publisher = {American Physical Society},
	doi = {10.1103/PhysRevD.102.014045}
}

@article{sizykh_2024_transmission,
	title = {Transmission of Vortex Electrons through a Solenoid},
	author = {Sizykh, G. K. and Chaikovskaia, A. D. and Grosman, D. V. and Pavlov, I. I. and Karlovets, D. V.},
	year = 2024,
	month = apr,
	journal = {Physical Review A},
	volume = {109},
	number = {4},
	pages = {L040201},
	doi = {10.1103/PhysRevA.109.L040201}
}

@misc{zmaga_2025_radiation,
	title = {Radiation of "Breathing" Vortex Electron Packets in Magnetic Field},
	author = {Zmaga, G. V. and Sizykh, G. K. and Grosman, D. V. and Meng, Qi and Zou, Liping and Zhang, Pengming and Karlovets, D. V.},
	year = 2025,
	month = dec,
	number = {arXiv:2509.21195},
	eprint = {2509.21195},
	primaryclass = {quant-ph},
	publisher = {arXiv},
	doi = {10.48550/arXiv.2509.21195},
	archiveprefix = {arXiv}
}

@article{maksimov_2025_diffraction,
	title = {Diffraction by Circular and Triangular Apertures as a Diagnostic Tool of Twisted Matter Waves},
	author = {Maksimov, M. and Borodin, N. and Kargina, D. and Naumov, D. and Karlovets, D.},
	year = 2025,
	month = dec,
	journal = {Physical Review A},
	volume = {112},
	number = {6},
	pages = {062823},
	doi = {10.1103/z2rs-2ryl}
}

@article{karlovets_2026_angular,
	title = {Angular Momentum Dynamics of Vortex Particles in Accelerators},
	author = {Karlovets, D. and Grosman, D. and Pavlov, I.},
	year = 2026,
	month = feb,
	journal = {Physical Review Letters},
	volume = {136},
	number = {8},
	pages = {085002},
	doi = {10.1103/gsrz-cscl}
}

@article{murtazin_2026_photon,
	title = {Photon Emission by Vortex Particles Accelerated in a Linac},
	author = {Murtazin, A. {\relax Yu}. and Sizykh, G. K. and Grosman, D. V. and Rybak, U. G. and Shchepkin, A. A. and Karlovets, D. V.},
	year = 2026,
	month = feb,
	journal = {Physical Review D},
	volume = {113},
	number = {3},
	pages = {036024},
	doi = {10.1103/g1xl-s2j5}
}

@misc{dyatlov_2026_classical,
	title = {Classical and Quantum Beam Dynamics Simulation of the {{RF}} Photoinjector Test Bench},
	author = {Dyatlov, A. S. and Kobets, V. V. and Levichev, A. E. and Maksimov, M. V. and Nikiforov, D. A. and Nozdrin, M. A. and Popov, K. and Sibiryakova, K. A. and Yunenko, K. E. and Karlovets, D. V.},
	year = 2026,
	month = feb,
	number = {arXiv:2509.00732},
	eprint = {2509.00732},
	primaryclass = {physics},
	publisher = {arXiv},
	doi = {10.48550/arXiv.2509.00732},
	archiveprefix = {arXiv}
}

@misc{dyatlov_2026_generation,
	title = {Generation of High-{{OAM}} Ultraviolet Twisted Light for {{RF-photoinjector}} Applications},
	author = {Dyatlov, A. S. and Dolgintsev, D. M. and Gerasimov, V. V. and Kobets, V. V. and Nazmov, V. P. and Nozdrin, M. A. and Sergeev, A. N. and Shokin, D. S. and Yunenko, K. E. and Karlovets, D. V.},
	year = 2026,
	month = feb,
	number = {arXiv:2512.08442},
	eprint = {2512.08442},
	primaryclass = {quant-ph},
	publisher = {arXiv},
	doi = {10.48550/arXiv.2512.08442},
	archiveprefix = {arXiv}
}

@article{lu_2006_nonlinear,
	title = {Nonlinear Theory for Relativistic Plasma Wakefields in the Blowout Regime},
	author = {Lu, W. and Huang, C. and Zhou, M. and Mori, W. B. and Katsouleas, T.},
	year = 2006,
	month = apr,
	journal = {Physical Review Letters},
	volume = {96},
	number = {16},
	pages = {165002},
	doi = {10.1103/PhysRevLett.96.165002}
}

@article{esarey_2009_physics,
	title = {Physics of Laser-Driven Plasma-Based Electron Accelerators},
	author = {Esarey, E. and Schroeder, C. B. and Leemans, W. P.},
	year = 2009,
	month = aug,
	journal = {Reviews of Modern Physics},
	volume = {81},
	number = {3},
	pages = {1229--1285},
	doi = {10.1103/RevModPhys.81.1229}
}

@article{tajima_2020_wakefield,
	title = {Wakefield Acceleration},
	author = {Tajima, T. and Yan, X. Q. and Ebisuzaki, T.},
	year = 2020,
	month = may,
	journal = {Reviews of Modern Plasma Physics},
	volume = {4},
	number = {1},
	pages = {1--72},
	doi = {10.1007/s41614-020-0043-z},
	copyright = {2020 The Author(s)}
}

@misc{lindstrom_2025_beamdriven,
	title = {Beam-Driven Plasma-Wakefield Acceleration},
	author = {Lindstr{\o}m, C. A. and Corde, S. and D'Arcy, R. and Gessner, S. and Gilljohann, M. and Hogan, M. J. and Osterhoff, J.},
	year = 2025,
	month = apr,
	number = {arXiv:2504.05558},
	eprint = {2504.05558},
	primaryclass = {physics},
	publisher = {arXiv},
	doi = {10.48550/arXiv.2504.05558},
	archiveprefix = {arXiv}
}

@article{manwani_2025_analysis,
	title = {Analysis of the Blowout Plasma Wakefields Produced by Drive Beams with Elliptical Symmetry},
	author = {Manwani, P. and Kang, Y. and Mann, J. and Naranjo, B. and Andonian, G. and Rosenzweig, J. B.},
	year = 2025,
	month = aug,
	journal = {Physical Review Letters},
	volume = {135},
	number = {9},
	pages = {095001},
	doi = {10.1103/28c4-blhg}
}

@article{ha_2022_bunch,
	title = {Bunch Shaping in Electron Linear Accelerators},
	author = {Ha, G. and Kim, K.-J. and Power, J. G. and Sun, Y. and Piot, P.},
	year = 2022,
	month = may,
	journal = {Reviews of Modern Physics},
	volume = {94},
	number = {2},
	pages = {025006},
	doi = {10.1103/RevModPhys.94.025006}
}



\end{document}